\documentclass[preprint, 12pt,authoryear]{elsarticle}

\usepackage{amssymb}
\usepackage{amsmath}
\usepackage{url}
\usepackage{xcolor} 
\definecolor{darkblue}{RGB}{0,0,139} 
\definecolor{gray}{RGB}{128,128,128} 

\usepackage[capitalise]{cleveref}
\crefname{figure}{Figure}{Figures}

\usepackage{enumitem} 
\newcommand*{\challthree}{Challenge 3: Real-world environments and dynamics compromise sensor reliability.}
\newcommand*{\challtwo}{Challenge 2: IMUs do not measure their orientations.}
\newcommand*{\challone}{Challenge 1: IMUs do not provide direct information about the joint angle.}
  
\newenvironment{longdescription}
  {\begin{description}[style=unboxed]}
  {\end{description}}

\journal{Journal of Biomechanics}

\begin{document}

\begin{frontmatter}



\title{Inertial Human Motion Capture: From Biomechanics to Recent Sensor Fusion Methods and Back} 

\author{Manon Kok$^\star$, Ive Weygers$^\dagger$, Hassan Osman$^{\star \star}$, Daniel Weber$^{\dagger \dagger}$, \\ Ruiyuan Li$^\star$, Thomas Seel$^{\dagger \dagger}$ and Ajay Seth$^{\star \star}$}

\affiliation{organization={$^\star$Delft Center for Systems and Control, Delft University of Technology},
            country={the Netherlands}}
\affiliation{organization={$^\dagger$Interuniversity Microelectronics Centre (imec)},
            country={Belgium}}            
\affiliation{organization={$^{\star \star}$Department of Biomechanical Engineering, Delft University of Technology},
            country={the Netherlands}}
\affiliation{organization={$^{\dagger \dagger}$Institute of Mechatronic Systems, Leibniz University Hannover},
            country={Germany}}            

\begin{abstract}
Inertial measurement units (IMUs) are a promising means to capture human motion, yet obtaining meaningful biomechanical quantities from IMU measurements remains non-trivial. This tutorial-style review focuses on kinematics and introduces four key aspects (inertial human motion capture objective, environmental conditions, subject \& attributes, and motion characteristics) to determine how to translate biomechanical problems into adequate formulations for the fusion of inertial sensor measurements. We identify three fundamental challenges for kinematics estimation from IMUs: IMUs do not provide direct information about the joint angle, IMUs do not measure their own orientation, and real-world environments and dynamics compromise sensor reliability. Though there exist widely-used methods to overcome these challenges, they suffer from severe limitations in real-life applications, e.g., the need for sensor-to-segment calibration, and the fact that magnetic field disturbances degrade joint angle accuracy. The full potential for many use-cases hence remains untapped in terms of accuracy and reliability. We share insights into recently proposed methods, e.g.\ exploiting the human body's kinematic chain constraints, having the potential to overcome these limitations. We also present guiding questions related to the four key aspects and illustrate their use for navigating the methodological landscape for the use-case of lower-extremity joint angle estimation, for which we share open-access code and compare the traditional workflow with three alternatives. Our aim is to bridge the gap between the sensor fusion community developing methods for human motion capture and the biomechanics community in need of accurate, easy-to-use, and reliable methods to study human motion outside of the laboratory.
\end{abstract}




\end{frontmatter}



\section{Introduction}
\label{sec:introduction}
Measuring human motion has important applications in entertainment, sports, ergonomics, and healthcare~\citep{suoTL:2024,liaoVXPB:2020,bastenJK:2009,menache:2000}. Particularly in healthcare, human motion capture has the potential to improve the quality of life of individuals with movement disorders~\citep{salisuRENSY:2023,knippenbergVLPTS:2017,lamTF:2023,mohrFPSAW:2023,hilderninkEtAl:2020}. These movement disorders also trigger secondary effects, including falls, obesity, and cardiovascular diseases that drive healthcare costs worldwide~\citep{nguyenLRCFN:2025}. The prevalence and secondary effects of movement disorders underscore the urgent need for accurate, easy-to-use tools to assess human movement.

Although human movement can be assessed using traditional optical motion capture systems, these have inherent shortcomings. First, individuals are known to behave differently in laboratory settings compared to their natural environments~\citep{roblesEtAl:2015,wadeNMB:2022}. Furthermore, these systems require trained technicians and substantial investments in infrastructure and equipment. These shortcomings restrict widespread clinical adoption of motion capture in healthcare. Inertial measurement units (IMUs) present a compelling alternative as small, portable, and inexpensive sensors that enable kinematic measures outside controlled laboratory environments, reducing costs and dependence on specialized infrastructure and expertise~\citep{priscoPSECAD:2025,renggliEtAl:2020}. Accessibility of IMU systems could democratize biomechanical analysis across diverse healthcare and human performance settings.

Biomechanical analysis requires accurate quantification of human motions in terms of joint angles, i.e.\ the angles between segments of the body that describe how segments move relative to one another~\citep{winter:2009}. Obtaining joint angles from IMU measurements is the focus of this paper in which we:
\begin{itemize}
    \item Present a tutorial-style review aimed at bridging the gap between the sensor fusion community developing methods for human motion capture and the biomechanics community. To this end, we introduce four key aspects that collectively determine how to translate biomechanical problems into adequate formulations for the fusion of inertial sensor measurements. 
    \item Identify three fundamental challenges to derive joint angles from inertial sensors and critically discuss inherent limitations of widely-used methods to overcome these challenges, e.g., the need for sensor-to-segment calibration, and the fact that the presence of magnetic field disturbances degrade the joint angle accuracy. 
    \item Share insights into recent state-of-the-art advancements, demonstrating how modern tools overcome traditional limitations (mostly by exploiting the human body's kinematic chain constraints), thereby improving the reliability and ease-of-use of joint angle estimation. 
    \item Present application-agnostic key study-design questions to guide the biomechanics community on how and when to use which method to obtain joint angles from IMU measurements.
    \item Exemplify the entire workflow on a representative real-world human movement dataset, providing fully open-access code and data\footnote{\url{https://github.com/daniel-om-weber/imu-joint-angle-methods-supplement}} to maximize reproducibility.
\end{itemize}

We hope that this work will enable biomechanical researchers to exploit the current breadth of available IMU-based methods to achieve accuracy and reliability sufficient for longitudinal clinical deployment, and that it will serve as inspiration for the sensor fusion community to tackle remaining open challenges. Our contribution contrasts with previous reviews that have focused on specific extremities or applications~\citep{chenLLY:2016,liQGC:2024,leeL:2022,guLHZZ:2023,chambersGCB:2015}.


\section{From IMU measurements to human motion capture use-cases}
\label{sec:hmc-problem}
\begin{figure}[htbp]
  \centering
\includegraphics[width=\linewidth]{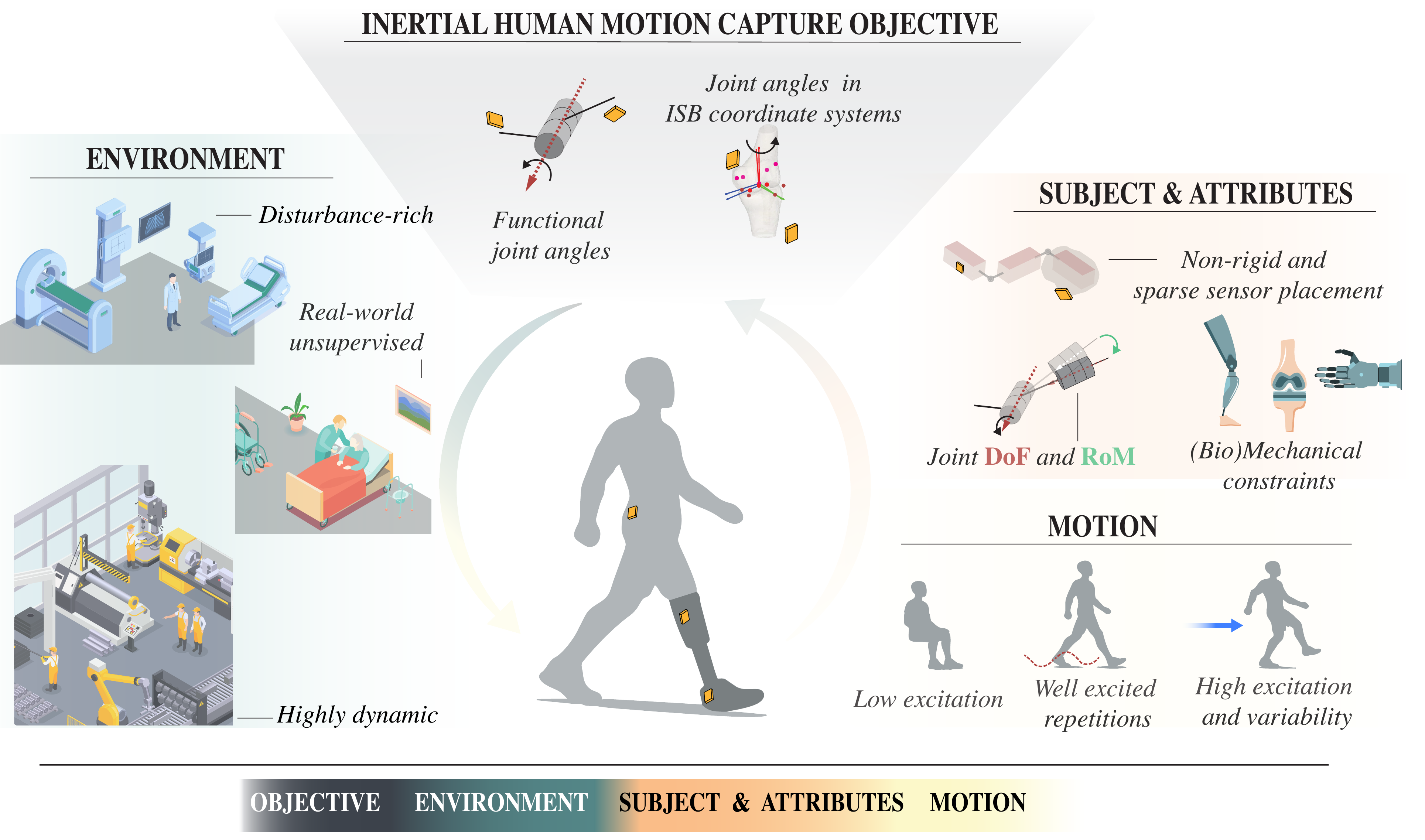}
  \caption{Four aspects, including the specific \emph{inertial human motion capture objective}, conditions of the \emph{environment}, \emph{subject and attributes}, and \emph{motion} dynamics, collectively determine how biomechanical problems are translated into formulations for the fusion of inertial sensor measurements. A thorough understanding of the influence of each aspect on accuracy and reliability is essential for designing clinically relevant and effective assessments based on IMUs.}
  \label{fig:isf-problem}
\end{figure}

Obtaining joint angles from IMU measurements requires fusing the available sensor measurements, which is also called sensor fusion. The accuracy and reliability of the obtained joint angles depends on multiple interacting factors. More insight into these factors can be obtained by studying the sensor fusion formulations. These formulations are typically application-agnostic and therefore applicable to a wide variety of biomechanical use-cases, body segments, and diverse musculoskeletal and neurological conditions. They are governed by e.g.\ movement assumptions and joint constraints, and critically depend on (i) the specific \emph{inertial human motion capture objective}, (ii) \emph{environment} in which the motion occurs, (iii) the \emph{subject and their attributes}, and (iv) the characteristics of the \emph{motion} (Figure \ref{fig:isf-problem}).
    
Throughout this paper, we illustrate how joint angles can be obtained from IMU measurements with the following use-case: early identification of abnormalities in gait patterns after a clinical intervention. An important functional indicator can be the kinematic changes in the lower limbs, which are ideally recorded within a few days after the intervention. 

Several aspects of the inertial motion capture task underlie this use-case. The first aspect is a critical reflection on the \emph{inertial human motion capture objective} in terms of the quantity of interest, e.g. which joint angles, and their required accuracy for the intended use-case. Second, the need for early identification highlights the influence of the \emph{environment}, as this typically implies that this needs to be done in indoor clinical settings, which are disturbance-rich environments where equipment and people are constantly moving. Third, injuries and interventions often change the \emph{subject}'s structure and biomechanical \emph{motion}. This change should be taken into account, as it may violate assumptions made by the sensor fusion formulations, and potentially restrict both the number of sensors and their placement on the subject. 

Building on these insights, Figure \ref{fig:isf-problem} provides a holistic view of the aspects shaping inertial human motion capture approaches for biomechanical problems requiring joint angle information. This perspective is intended to support biomechanical researchers in navigating the associated challenges and limitations of inertial human motion capture (Section \ref{sec:hmc_limitations}), and position them to critically assess recent algorithmic developments outlined in Section~\ref{sec:overcoming}.


\section{From IMUs to joint angles: Challenges and limitations}
\label{sec:hmc_limitations}
There are three inherent challenges to obtain joint angles from IMUs. The most commonly-used workflow to overcomes these challenges, see e.g.~\citet{albornoEtAl:2022}, is visualized in \cref{fig:overview-section3}. Here, IMU measurements are used to estimate sensor orientations. Sensor-to-segment assignment is used to determine which sensor is placed on which body segment. Furthermore, sensor-to-segment calibration is used to determine the placement of the sensors and their alignment to the body segments and respective joint axes. This commonly-used workflow has inherent limitations related to the accuracy of the obtained joint angles and to its ease-of-use. These limitations are discussed in \cref{sec:hmc_limitations_summary}. 

\subsection{Sensor orientations vs human joint angles: Calibration and assignment}
\label{sec:hmc-using-imus}
\begin{figure}
  \centering
  \includegraphics[width=\linewidth]{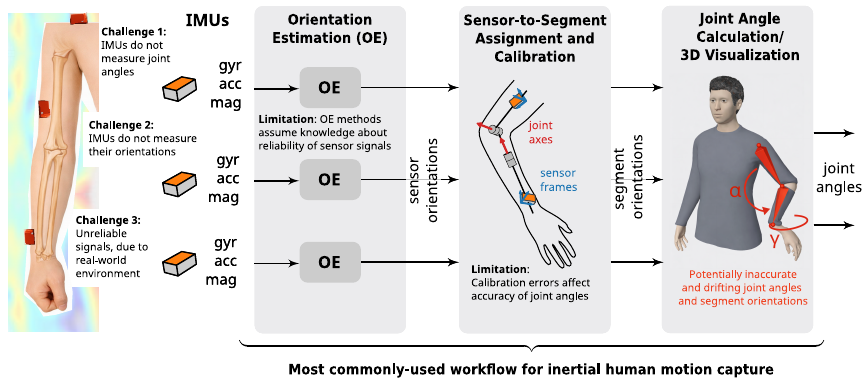}
  \caption{An overview of the most commonly-used workflow to obtain joint angles from IMU measurements, including a summary of the challenges and limitations.}
  \label{fig:overview-section3}
\end{figure}

Human motion capture can be abstracted to the problem of estimating relative angles of adjacent segments connected by joints in a kinematic chain. To estimate the motion of this kinematic chain, it is common practice to place an IMU per body segment comprising such a kinematic chain, see also \cref{fig:overview-section3}. While IMU sensor orientations provide important information for human motion capture, they specifically \emph{do not} directly capture human joint angles, which brings us to the first challenge of inertial human motion capture:

\begin{longdescription}
\item[\challone] They instead provide information about the movement of the sensor and do not directly relate to movement of the body or to anatomical frames. 
\end{longdescription}

To formalize Challenge 1, let us, for simplicity, assume that each IMU outputs its own \emph{absolute orientation} (see also \cref{sec:imuOri}). This absolute orientation is defined as the orientation between the \emph{sensor frame} of this IMU and a \emph{global inertial frame}, which is typically aligned with the directions of gravity and local magnetic north. The \emph{relative orientation between two IMUs} can straightforwardly be computed from the absolute orientation of these IMUs. This relative orientation varies with time when the human body moves, and therefore provides information about the movement of the human body. This will, however, be expressed in terms of rotations around sensor axes and will therefore depend on how the sensor was mounted on the body segment. Hence, IMUs do not directly provide information about the joint angles, which are defined by rotations about specific axes defined by the anatomical coordinate system \citep{wuEtAl:2002,wuEtAl:2005}.

It is possible to determine the joint angle from the IMU orientations when the mounting of the sensors on the body segments is known or has previously been determined. The commonly-used workflow for determining this mounting consists of two steps. First, sensor-to-segment assignment specifies which sensor is attached to which body segment. This is typically done manually, see e.g.\ \cite{cuttiFGRCF:2010}. Second, sensor-to-segment calibration determines the placement of the sensors and their alignment to the body segments / joint axes. This sensor-to-segment calibration is typically done by manual alignment, static pose calibration, functional calibration, anatomical landmark calibration, or combinations thereof \citep{ekdahlLESU:2023,pacherCVF:2020}. Each of these sensor-to-segment calibration methods implicitly includes information about the body segments and their relative movements about joints, thereby allowing for mapping the sensor orientations to segment orientations and subsequently to joint angles, see \cref{fig:overview-section3}. These can be represented in terms of joint angles according to the recommendations of the International Society of
Biomechanics (ISB) \citep{wuEtAl:2002,wuEtAl:2005,cereattiEtAl:2024} or in terms of functional joint angles (i.e., about  axes with a direction that is derived from motions rather than bony landmarks \citep{ehrigTDH:2007}). They are often computed using inverse kinematics methods, see e.g. \citet{sethEtAl:2018}. 

\subsection{IMU measurements vs sensor orientations: Orientation estimation}
\label{sec:imuOri}
Although many IMUs output sensor orientations, as we assumed in \cref{sec:hmc-using-imus}, they \emph{do not} directly measure their own orientation. Instead, they consist of a 3-axis accelerometer, a 3-axis gyroscope, and often a 3-axis magnetometer. For each IMU, these measurements can be combined to estimate the absolute orientation of the sensor. Note that whenever we refer to orientation estimation and to sensor orientations, we implicitly refer to absolute orientations between the sensor frame and a global inertial frame aligned with the directions of gravity and local magnetic north.

\begin{longdescription}
\item[\challtwo] Instead, sensor fusion is needed to combine measurements from the gyroscope, accelerometer, and magnetometer. 
\end{longdescription}

The field of sensor fusion combines (imperfect) sensor measurements and models to estimate quantities of interest, in this case sensor orientations. The measurements from the three sensors in an IMU can be used in these sensor fusion formulations and provide complementary information about the sensor's orientation as~\citep{kokHS:2017}:
\begin{description}
\item[\textit{Gyroscope}] The angular velocity measured by the gyroscope can be summed over time to provide information about the change in orientation. The reliability of this information is sensor-dependent, with errors due to gyroscope bias and noise accumulating over time.
\item[\textit{Accelerometer}] The specific force measured by the accelerometer consists of the gravity vector measured in the sensor frame, and the linear acceleration of the sensor. Hence, when the linear acceleration is small compared to the magnitude of the gravity, the accelerometer can be used to provide information about the inclination (tilt) of the sensor, i.e.\ the angle between the sensor axes and the gravity vector. The reliability of this information is dynamics-dependent, since in highly dynamic movement the gravity
direction cannot be reliably isolated.
\item[\textit{Magnetometer}] The local magnetic field measured by the magnetometer consists of the local Earth's magnetic field and any magnetic field anomalies. Hence, when the magnetic field anomalies are small compared to the magnitude of the local Earth's magnetic field, the magnetometer provides information about the angle between the sensor axes and the local Earth's magnetic field vector. Unless the sensor is on one of the Earth's poles, this results in heading information. The reliability of this information is environment-dependent, since in magnetically disturbed environments the Earth's magnetic field vector cannot be reliably isolated.
\end{description}
These sensors therefore provide full 3D absolute orientation information only when combined, where some of the information is redundant. As an example, when we would obtain an initial orientation using accelerometer and magnetometer measurements, the subsequent orientations can be estimated using gyroscope measurements only. However, these estimates will drift due to the accumulation of errors due to gyroscope bias and noise. On the other hand, when accelerometer and magnetometer measurements are used at every time instance, this leads to estimation errors in the presence of linear accelerations and magnetic field anomalies. Any method for orientation estimation, therefore, needs to encode knowledge about the reliability of each sensor's measurement. This reliability cannot be quantified a priori because it depends on the environment and sensor dynamics. This leads us to the third challenge: 

\begin{longdescription}
\item[\challthree] Specifically, magnetometer measurements in practice often provide unreliable heading information due to the presence of magnetic field anomalies in real-world environments. Furthermore, the inclination information from the accelerometer degrades under real-world dynamics, i.e. in the presence of linear accelerations. How much the orientation estimates degrade depends on the settings and models used in the sensor fusion methods. 
\end{longdescription}

Non-zero linear acceleration of the sensor is of particular concern for dynamic tasks~\citep{luingeV:2004}. In practice, however, this issue is often less severe than the magnetic field anomalies. This is because the acceleration due to gravity is often large compared to human accelerations and even in dynamic tasks there are typically specific time instances with less acceleration that can be used for inclination information. Magnetic field anomalies on the other hand, are typically present in all indoor environments as they are caused by ferromagnetic material in the vicinity of the sensors, such as steel in the construction of buildings \citep{liGDR:2012,deVriesVBH:2009}. These magnetic field anomalies vary over space, and are often more severe closer to the ground. A magnetic field that varies over space is also visualized in \cref{fig:overview-section3}.  

Commonly-used methods for orientation estimation, such as \citet{madgwickHV:2011} and \citet{mahonyHP:2008}, use a single, constant parameter to model the reliability of the sensor measurements. This parameter will by definition not be the ``optimal'' parameter at almost all time instances \citep{caruso:2021}. Because of this, obtaining accurate, non-drifting sensor orientations still remains a challenge. Furthermore, proprietory orientation estimation algorithms used in commercially available IMUs typically lack transparency on the exact parameter and model choices, posing limitions also for validation and reproducibility. This is of even bigger concern for human motion capture than for most other applications, since unreliability of the measurements of one sensor will cause the human motion estimates to become inconsistent and potentially physically infeasible.

\subsection{Limitations of commonly-used workflow to obtain joint angles from IMUs}
\label{sec:hmc_limitations_summary}
In this section, we have discussed the common workflow for inertial human motion capture and the challenges in obtaining joint angles from IMU measurements. In \cref{sec:hmc-using-imus}, we introduced the challenge that IMUs do not measure joint angles. Although there exist widely-used methods to obtain joint angles from IMU orientations, they face several limitations: 
\begin{description}
\item[Limitation 1:] Calibration errors directly affect the quality of the joint angle estimates. Furthermore, calibration implies additional setup time for the system.
\item[Limitation 2:] The location of the sensor on the body is assumed constant over time, which in practice is not the case due to soft tissue artefacts. Furthermore, sensor-to-segment assignment is assumed to be known, which may add to the setup time.
\end{description}

A second and third challenge are concerned with IMUs not directly measuring orientations, and with real-world environments and dynamics compromising sensor reliability. This results in a loss of accuracy for orientation estimates and the following limitation:
\begin{description}
\item[Limitation 3] Sensor measurements in practice \emph{never} fulfill the underlying assumptions required by absolute orientation estimation methods. In other words, sensor biases are in practice not zero nor constant, magnetic field anomalies are present, and in any relevant scenario, the sensor's linear acceleration is non-zero, which leads to inaccuracies in estimated orientations.
\end{description}
In \cref{sec:overcoming} we will discuss new methods that have specifically been developed to overcome these limitations.


\section{Overcoming the limitations of the common inertial motion capture workflow}
\label{sec:overcoming}
\begin{figure}
  \centering
  \includegraphics[width=\linewidth]{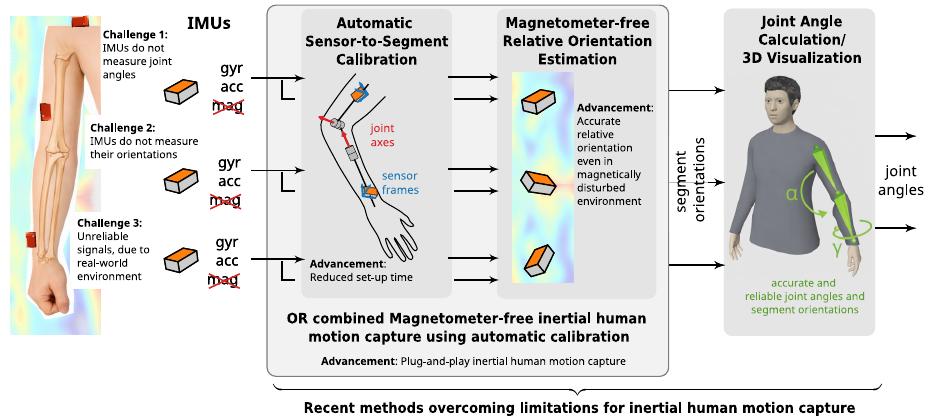}
  \caption{Recent sensor fusion methods to overcome the limitations of the common workflow from \cref{fig:overview-section3}.}
  \label{fig:overview-section4}
\end{figure}

Recent years have seen numerous developments in the field of sensor fusion to overcome the three limitations of the commonly-used workflow for inertial human motion capture introduced in \cref{sec:hmc_limitations_summary}. In this section, we categorize these recent methods into four classes. The first class of methods focuses on improving single-sensor (absolute) orientation estimates, while leaving the rest of the workflow from \cref{fig:overview-section3} untouched. The other three classes are visualized in \cref{fig:overview-section4}. They focus on magnetometer-free relative orientation estimation, on automatic sensor-to-segment calibration, and on the combined magnetometer-free inertial human motion capture using automatic calibration. 

\subsection{Improving single-sensor orientation estimates}
\label{sec:overcoming-single-sensor}
In recent years, several methods have been developed that focus on minimizing the negative effect of real-world environments and dynamics on the orientation estimates (see Challenge 3). In other words, they focus on improving orientation estimates in the common workflow visualized in \cref{fig:overview-section3}, while leaving the rest of the workflow untouched. To improve the orientation estimates, these methods either adapt the model assumptions from the commonly-used absolute orientation estimation methods to ensure that the sensor measurements are less likely to violate their assumptions (directly focusing on overcoming Limitation 3), or they use machine learning. Machine learning has, for instance, been used to directly learn orientations from IMU measurements \citep{weberGS:2021,esfahaniWWY:2019}. It has also been used to learn e.g.\ inertial sensor biases and noises \citep{brossardBB:2020,engelsmanK:2022}.

Orientation estimation methods adapting model assumptions have mainly focused on adapting the model such that the negative influence of magnetic field disturbances on the orientation estimates is mitigated. Specifically, since the magnetometer provides information about the angle between the sensor axes and the local Earth's magnetic field, it not only provides information about the sensor's heading but also about its inclination (except on the equator). Since the accelerometer typically provides more reliable inclination information, these works focus on more effectively allowing the magnetometer to only influence the heading information \citep{seelR:2017,laidigS:2023} than the widely-used method from \citet{madgwickHV:2011}. Combining this with adapted models that mitigate the influence of temporal accelerations and magnetic field anomalies has been shown to outperform a wide range of widely-used algorithms on a wide variety of datasets \citep{laidigS:2023}. 

\subsection{Magnetometer-free relative orientation estimation}
\label{sec:overcoming-magfree}
While the methods from \cref{sec:overcoming-single-sensor} make important steps towards overcoming Limitation 3, magnetic field anomalies are so common in indoor environments --- as they are caused e.g.\ by reinforced steel in the construction of buildings --- that there is also a large body of recent literature that focuses on \emph{magnetometer-free} inertial human motion capture. While the methods from \cref{sec:overcoming-single-sensor} estimate \emph{absolute orientation}, still needing to rely on magnetometer measurements to mitigate drift in the heading estimates, the magnetometer-free methods that we will discuss in this section estimate the \emph{relative orientation} between sensors on adjacent body segments. Note that the relative orientation between the sensors is sufficient to obtain joint angles, see also the workflow in \cref{fig:overview-section4}. 

The relative orientation can be estimated without the use of magnetometers when information about the kinematic chain is known and included in the sensor fusion problem. More specifically, let us assume that we are interested in estimating the joint angles between a number of connected body segments using an IMU on each segment. The relative orientation between these sensors can be estimated using only accelerometer and gyroscope measurements by utilizing the knowledge that the body segments are connected. The very mild conditions on the motion of the segments that is required to be able to estimate the relative orientation between the sensors --- the most notable one that the body segments should not be completely stationary --- have been mathematically derived and experimentally tested in \cite{kokEWS:2022}. Magnetometer-free relative orientation estimation has been considered for 1D and 2D joints, see e.g.\ \citet{lehmannLBSW:2024,lehmannLDS:2020,truppaBGVSM:2022,seelS:2013}, and general 3D joints \citep{weygersKVVHVC:2020,kokHS:2014,leeJ:2019,lehmannLDS:2020,truppaBGVSM:2022}.

These methods for magnetomer-free relative orientation estimation require slightly different processing and practical considerations than the common workflow from \cref{sec:hmc_limitations}. First, it is no longer possible to process the measurements from each sensor separately, as estimating relative orientation requires joint processing of the measurements from multiple sensors. Furthermore, while non-zero linear accelerations degrade the performance of the absolute orientation estimation methods from \cref{sec:hmc_limitations}, magnetometer-free relative estimation methods actually require motion to obtain accurate relative orientation. For methods considering 1D or 2D joints, an additional requirement is that the joint axis is not (nearly) vertical for extended periods of time as in this case the accelerometers on the adjacent segments do not provide relative heading information. Existing work has explored mitigating these requirements by additionally estimating a heading offset \citep{lehmannLDS:2020}.

\subsection{Automatic sensor-to-segment calibration}
\label{sec:overcoming-i2scal}
In recent years, automatic calibration methods to determine the placement of sensors on segments have been developed. These automatic sensor-to-segment calibration methods replace the manual alignment methods in the common workflow, see \cref{fig:overview-section3,fig:overview-section4}. They also do not rely on the user performing specific movements. Instead, the methods automatically determine the sensor-to-segment calibration based on normal human motion, allowing for plug-and-play use. Hence, they overcome Limitation 1, reducing the set-up time of the system, and avoiding errors introduced during manual alignment. Similar to the methods from \cref{sec:overcoming-magfree}, these automatic sensor-to-segment calibration methods make use of information about the connectivity of the body segments and potentially about the limited degrees-of-freedom of the joints \citep{seelSR:2012,olssonKSH:2020}. 

\subsection{Magnetometer-free inertial human motion capture using automatic calibration}
\label{sec:overcoming-calfree}
The automatic sensor-to-segment calibration techniques from \cref{sec:overcoming-i2scal} and the fact that these techniques use similar models as the magnetometer-free methods from \cref{sec:overcoming-magfree}, opens up for combining these, overcoming both Limitation 1 and Limitation 3 of the methods from \cref{sec:hmc_limitations} in a single algorithm. 

Magnetometer-free inertial human motion capture using automatic calibration is specifically well-developed for the case where the joint is assumed to be 1D or 2D. In this case, the automatic calibration determines the direction of the joint axes in the sensor frame and directly aids the computation of the joint angles. For general 3D joints, magnetometer-free human motion capture using automatic calibration has recently been proposed in \cite{taetzLMSB:2025,lorenzTBS:2025}. Note that in this case, the automatic calibration determines the location of the sensors on the body. In 3D joint models, joint axes are inherently not uniquely defined. Hence, to obtain joint angles (see the last step in \cref{fig:overview-section4}), joint angle projection is still needed, and may nevertheless require additional calibration. A fully plug-and-play method for 3D joints is therefore not yet available. Although these solutions for 3D joints are not available as open source tools, we expect that new methodological advances will result in open solutions in the near future. 

\section{From guiding questions to method selection}
\label{sec:use-case-comparison}

\begin{figure}[htbp]
  \centering
  \includegraphics[width=\linewidth]{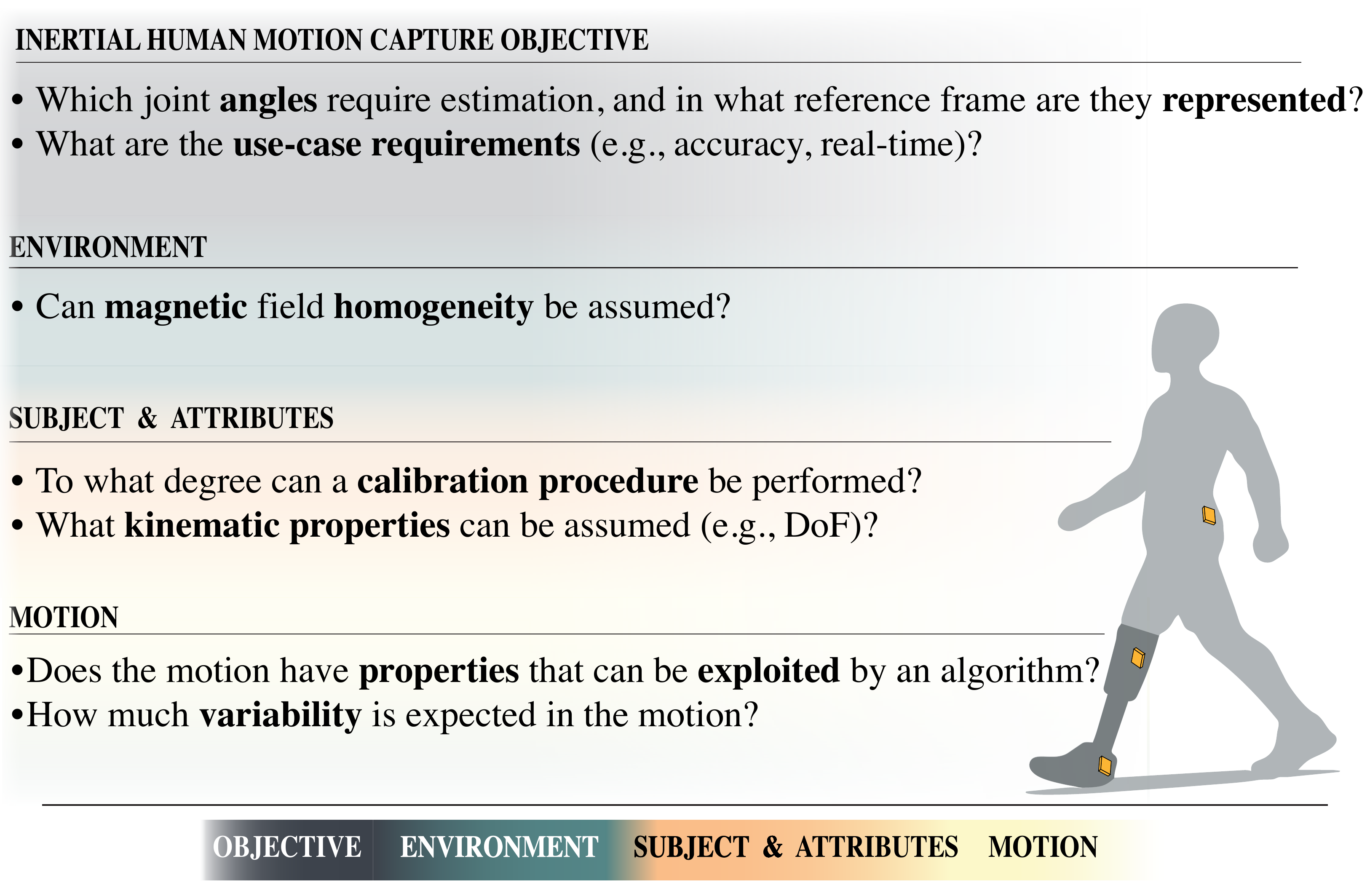}
  \caption{Principled questions guiding the translation of a biomechanical application into a sensor fusion formulation, covering the estimation objective, environmental conditions, motion characteristics, and the subject of study and its attributes.}
  \label{fig:isf-q}
\end{figure}

We present a set of principled questions (\cref{fig:isf-q}) to guide the selection of an estimation method (\cref{sec:hmc_limitations,sec:overcoming}) for extracting joint angles in a specific biomechanical application. These guiding questions are directly related to the four key aspects from \cref{fig:isf-problem}. We demonstrate how these questions can guide the selection of an estimation method on the use-case of early identification of abnormalities in gait patterns after a clinical intervention, introduced in \cref{sec:hmc-problem}. We compare four processing strategies (\emph{Madgwick + IK}, \emph{VQF + IK}, \emph{VQF + Olsson}, and \emph{Weygers}), progressing from the common workflow (\cref{fig:overview-section3}) to recent magnetometer-free sensor fusion methods (\cref{fig:overview-section4}). Each strategy assembles building blocks from \cref{sec:hmc_limitations,sec:overcoming} into a complete processing pipeline for which we made open-source code available.\footnote{\url{https://github.com/daniel-om-weber/imu-joint-angle-methods-supplement}} Although the lower extremities serve as the running example, these methods apply to the upper extremities as well. 

The guiding questions (\cref{fig:isf-q}) characterize this scenario as follows. The \emph{estimation objective} is to obtain knee and ankle joint angles defined according to ISB conventions, i.e.\ with respect to anatomically defined axes. Offline processing is acceptable. The \emph{environment} is an indoor hospital room where the magnetic field is inhomogeneous due to reinforced steel in the building and nearby medical equipment, but no major magnetic disturbances are present. The expected \emph{motion} is continuous level walking with repetitive gait cycles that provide sufficient excitation for the magnetometer-free relative orientation estimation methods (\cref{sec:overcoming-magfree}). For the \emph{subject \& attributes}, a sensor-to-segment calibration step is possible, with the sensors manually assigned and aligned. The kinematic model assumes the knee as a 1D hinge joint (dominant flexion/extension) and the ankle as a 2D joint (dorsiflexion/plantarflexion and inversion/eversion). These answers narrow the choice in two respects: the \emph{environment} decides whether the magnetometer can be trusted, separating the magnetometer-based from the magnetometer-free pipelines, while the joint type (\emph{subject \& attributes}) and the \emph{objective} together set the joint model, in particular whether one functional axis suffices for the 2D ankle. The four pipelines span this space, and the comparison relates each answer to the resulting accuracy.

We evaluate the four methods on healthy participants, isolating methodological differences from patient-specific confounds such as altered gait or restricted range of motion. We use data from four subjects performing 10-minute walking trials (5\,m straight, 180\textdegree{} turn, self-selected pace) from the OpenSense dataset \citep{albornoEtAl:2022}. Each subject wears eight Xsens MTw Awinda IMUs sampled at 100\,Hz, and the right lower extremity is evaluated. Ground-truth joint angles are obtained from optical motion capture processed with OpenSim inverse kinematics \citep{sethEtAl:2018} using a subject-scaled Rajagopal musculoskeletal model \citep{sethEtAl:2018, Rajagopal:2016}. Our comparison focuses on knee flexion/extension and ankle dorsiflexion/plantarflexion. Accuracy is quantified as the root-mean-square error (RMSE) per subject, averaged across the four subjects. For each individual use-case, a critical reflection of such RMSE values is always required to ensure that error ranges remain below clinically relevant task- and joint-dependend thresholds \citep{kuhlgatz2026inertial}. For instance, errors below 5 degrees are often considered acceptable for the lower extremities~\citep{mcginley2009reliability}.

\begin{figure}[htbp]
  \centering
  \includegraphics[width=\linewidth]{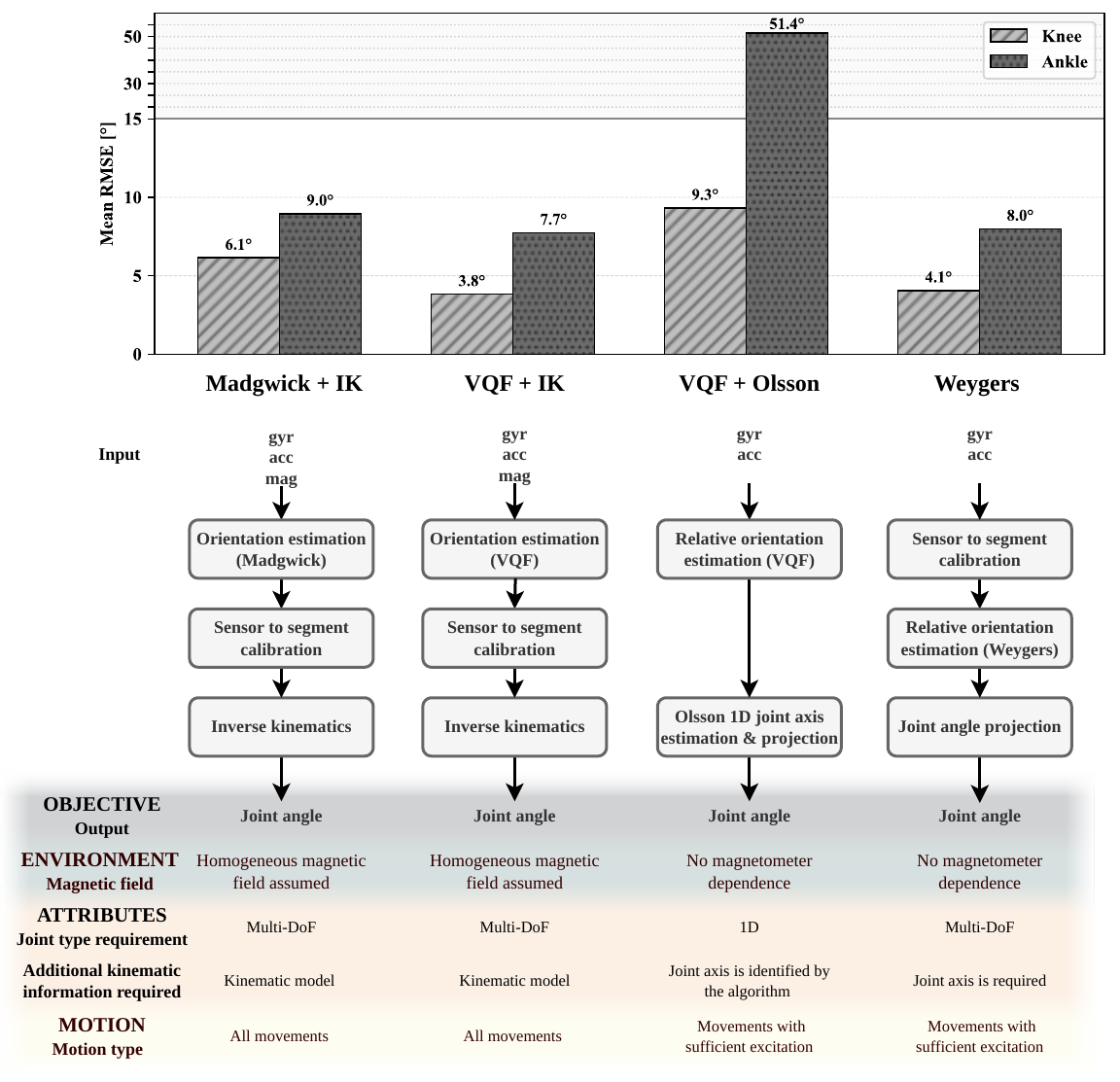}
  \caption{Comparison of four joint angle estimation pipelines evaluated on the OpenSense dataset \citep{albornoEtAl:2022}. Each column shows the processing steps from IMU data to joint angles, with rows below mapping the pipeline to the guiding questions from \cref{fig:isf-q} and the key aspects from \cref{fig:isf-problem}. RMSE values (mean across four subjects) are reported for knee flexion/extension and ankle dorsiflexion/plantarflexion.}
  \label{fig:use-case-comparison}
\end{figure}

The first approach (\emph{Madgwick + IK}) follows the common workflow from \cref{fig:overview-section3} and is not magnetometer-free. It estimates sensor orientations using the widely-used approach from \citet{madgwickHV:2011}, and computes joint angles via a kinematic model and inverse kinematics. This pipeline assumes a homogeneous magnetic field and a calibrated kinematic model. Because the Madgwick filter does not completely decouple magnetic corrections from inclination estimation (see also \cref{sec:overcoming-single-sensor}), these disturbances corrupt both heading and tilt, manifesting as jitter in the estimated joint angles. This results in an RMSE of 6.1\textdegree{} for the knee and 9.0\textdegree{} for the ankle.

The second approach (\emph{VQF + IK}) replaces the Madgwick filter with VQF \citep{laidigS:2023}, keeping the rest of the pipeline unchanged. This reduces the RMSE to 3.8\textdegree{} (knee) and 7.7\textdegree{} (ankle). This reduction arises because VQF more effectively decouples inclination estimation from magnetometer-based heading correction (\cref{sec:overcoming-single-sensor}), reducing the influence of magnetic disturbances on the inclination estimates. The method still uses the magnetometer for heading, whereas the remaining two approaches are magnetometer-free.

The third approach (\emph{VQF + Olsson}) eliminates the magnetometer and manual sensor-to-segment calibration (\cref{sec:overcoming-magfree,sec:overcoming-i2scal}), overcoming both Limitations~1 and~3. It uses VQF without magnetometers to estimate per-sensor absolute orientations, then computes relative orientations between adjacent segments. The method from \citet{olssonKSH:2020} then identifies the functional joint axis from the same measurements, assuming a 1D hinge joint. The relative orientation is projected onto this axis to obtain a single functional joint angle, bypassing inverse kinematics. For the knee, where flexion/extension dominates, the data-driven functional axis aligns well with the anatomical axis, yielding an RMSE of 9.3\textdegree{}, higher than the magnetometer-based pipelines due to residual heading drift. For the ankle, secondary motions, such as inversion and eversion, bias the functional axis identification, preventing alignment with the anatomical dorsiflexion/plantarflexion axis. The resulting functional angle mixes contributions from multiple anatomical directions, a phenomenon known as kinematic crosstalk \citep{Stephen:2000,lafortuneCSK:1992}, producing an unusable RMSE of 51.4\textdegree{}. Because the \emph{estimation objective} requires ISB angles along anatomical axes, this mismatch renders the method inapplicable to joints with two or more degrees of freedom.

Retaining magnetometer-free estimation while accommodating the ankle's multiple degrees of freedom requires a more general joint model, at the cost of reintroducing calibration. This fourth approach (\emph{Weygers}) has three stages. First, sensor-to-segment calibration: the sensor locations on the segments are automatically calibrated \citep{seelSR:2012} (\cref{sec:overcoming-i2scal}), while joint-axis identification is performed manually. Second, the magnetometer-free method from \citet{weygersKVVHVC:2020} estimates relative orientations between adjacent segments assuming a 3D joint (\cref{sec:overcoming-magfree}), a formulation valid for joints of any number of degrees of freedom. Third, the relative orientations are projected onto the identified anatomical axes to obtain ISB angles. The resulting RMSE (4.1\textdegree{} knee, 8.0\textdegree{} ankle) is similar to \emph{VQF + IK}, but achieved without magnetometers.

The comparison in \cref{fig:use-case-comparison} shows how the answers to the guiding questions (\cref{fig:isf-q}) guide the choice of pipeline. The near-tie between \emph{VQF + IK} and \emph{Weygers} is settled by the \emph{environment}: the mild disturbance here leaves the magnetometer usable, so \emph{VQF + IK} wins narrowly. Under stronger, persistent anomalies, a corrupted magnetometer heading propagates into the joint angle, so the magnetometer-free \emph{Weygers}, of comparable accuracy, becomes preferable where its manual joint-axis calibration is acceptable. Which pipeline performs best therefore depends on the \emph{environment} and joint type, so a ranking obtained in one setting does not transfer to another.


\section{Outlook}
\label{sec:outlook}
As illustrated in \cref{sec:use-case-comparison}, the methods discussed in \cref{sec:overcoming} hold great promise for biomechanical applications since they overcome Limitations 1 and 3 of current practice. It is, however, important to co-develop them further, where biomechanical researchers provide input from the use of the methods in real-life applications, and sensor fusion researchers explore methodological advances to overcome potential shortcomings. We expect, for instance, that new methods will be needed in use-cases of limited motion in highly disturbed magnetic environments, as limited motion reduces the accuracy of magnetometer-free approaches. Methodological developments may also be needed for use-cases with low computational budgets, with high impacts (causing spikes in the accelerometer measurements), or in the presence of large soft-tissue artefacts (see also Limitation 2). Furthermore, obtaining accurate uncertainty quantification of the estimated joint angles is a relatively unexplored research problem. Increasing the ease-of-use of the methods is another possible direction of future work. An example is to include automatic sensor-to-segment assignment (Limitation 2). Existing automatic sensor-to-segment assignment methods \citep{graurockSS:2016,weenkBBHV:2013,zimmermannTB:2018} are, to the best of the authors' knowledge, not available in open source, and future work could focus on developing magnetometer-free inertial human motion capture using automatic calibration and assignment.

In an ideal future, a plug-and-play sensor fusion method that is accurate and reliable in all biomechanical applications will be developed. In the meantime, we hope that this work helps navigate the existing available methods and understand their shortcomings and limitations. Noteworthy, when joint angle estimates are inaccurate, to determine the root cause it is essential to question all assumptions of the used methods, starting from the IMU signals. For example, is the magnetometer calibrated (i.e. is the magnitude of the signal constant when rotating the sensor), and are high impacts or high accelerations visible in the accelerometer signals? Subsequently, are the orientation estimates of the single sensors as expected, and are the orientation estimates of different sensors consistent with each other? Note that we expect that all currently-available IMUs can satisfactorily be used in all estimation methods discussed in this work and that hardware is not the bottleneck for obtaining accurate joint angles from IMUs. For a more detailed comparison, see e.g. \citet{pasciutoLBVSC:2015,caruso:2021,fanZCDLLS:2025}.

We also foresee methodological developments far beyond the currently available methods. One interesting direction is towards sparse sensor setups where only a small number of sensors are used to estimate the motion of the entire human body. \citet{eckhoffKSS:2020} has, for instance, shown that under mild conditions on the movement, it is possible to estimate the movement of three body segments that are connected by hinge joints using only two IMUs, placed on the outer segments. Furthermore, there is a significant and growing body of work that uses machine learning methods to enable full-body human motion capture using only a small number of (often six) inertial sensors \citep{huangKABHP:2018,wouweLFDL:2024,yi2022physical,marcardRBP:2017,bachhuberWLDS:2024}. These methods are currently mostly developed from the computer vision/computer graphics community, and their potential use within biomechanics to date remains underexplored. An important concern, however, is that these machine learning-based methods are strongly biased by the training data, which can be detrimental in clinical practice where the goal is to understand how far away from normal a patient might be. 

While the focus of this work has been on kinematics estimation, we see joint angles as a gateway into biomechanical analysis, for instance, opening opportunities to use inertial sensors for estimating human kinetics. Methods relying on similar assumptions as those from \cref{sec:overcoming} but using musculoskeletal models instead of the simple 1D / 2D / 3D joint models, have in fact already been developed and shown to enable both the estimation of kinematics and kinetics \citep{nitschkeDLEK:2024,osmanKBKS:2026,dorschkyNSBE:2019,haraguchiH:2024,koelewijnEtAl:2025}. The fact that these methods allow for rather straightforward inclusion of additional sensor modalities further increases their potential. 

\section{Discussion and Conclusions}
Our aim with this work was to bridge the gap between the sensor fusion community developing methods for human motion capture and the biomechanics community in need of easy-to-use, accurate, and reliable methods to study human motion outside of the laboratory. While the current commonly-used IMU-to-joint angles workflow has significant limitations --- such as sensitivity to magnetic field disturbances, linear accelerations, and reliance on sensor-to-segment calibration --- we have discussed recent methods overcoming these limitations as well as given an outlook of methodological development yet to come. We have presented application-agnostic key study-design questions to guide how and when to use which method to obtain joint angles from IMU measurements, and have illustrated on an open-source dataset using open-source code that addressing such questions will allow for joint angle estimation errors limited to a few degrees even in the presence of magnetic field disturbances.

For biomechanics applications in which the most commonly-used workflow represented in \cref{fig:overview-section3} is used, we recommend replacing traditionally popular approaches for absolute orientation estimation, such as \citet{madgwickHV:2011,mahonyHP:2008}, with recently developed methods for orientation estimation that have been shown to result in more accurate and reliable estimates. Specifically for the method from \cite{laidigS:2023}, a direct replacement seems straightforward and beneficial, since 1) the method has been shown to result in more accurate orientation estimates than existing methods on a wide range of datasets, 2) its computational complexity is in the same order of magnitude as the traditionally popular methods, and 3) the method decouples the influence of violating individual sensor modality assumptions while at the same time being based on the same principles as traditionally used.

For magnetically-disturbed environments, we recommend exploring the use of magnetometer-free relative orientation estimation, see e.g.\ \cite{weygersKVVHVC:2020}, which can also be combined with automatic calibration, see e.g.\ \cite{olssonKSH:2020}. The limitation of these methods is that they require motion, and that for the methods assuming 1D / 2D joints, the joint axes cannot be vertical for extended periods of time \cite{lehmannLBSW:2024}. We therefore recommend to use magnetometer-free methods with care if these requirements are not fulfilled. Very recent work has also explored either combining models of different joint types \cite{lehmannLBSW:2024} or combining kinematic models with magnetic field measurements \cite{skovWOHLD:2026}. 

While our focus in this paper was on estimating joint angles, it is important to recognize that joint angles are not always the primary variables of interest. Metrics such as step length, center-of-mass velocity, or acceleration may be more relevant in certain clinical or performance contexts. In such cases, general joint-angle-based pipelines may be suboptimal, as these quantities can often be estimated more directly from the inertial measurements.

In conclusion, we have demonstrated that many of the long-standing challenges and limitations in the field of inertial human motion capture become manageable once the measurement problem is approached in a structured manner. By explicitly considering only a handful of key aspects (the inertial human motion capture objective, environmental conditions, subject \& attributes, and motion characteristics), biomechanics researchers can more easily navigate the methodological landscape and select methods that meet their needs. This yields significant gains in accuracy and reliability, thanks to recent, directly applicable openly available methods for absolute and relative, magnetometer-free orientation estimation, and for automatic calibration. Looking ahead, we hope that this will be a source of inspiration for synergies: sensor fusion methods shaped by biomechanical insight, and biomechanical analyses based on what the measurements can actually reveal.

\section{Acknowledgements}
This work was supported by the Convergence Human Mobility Center, a flagship initiative funded by the Convergence Alliance (TU Delft,
Erasmus MC, Erasmus University Rotterdam). It was also supported by the Sensor AI Lab through the AI Labs Program of the Delft University of Technology.

During the preparation of this work, the author(s) used ChatGPT, CoPilot, and Claude in order to check the grammar and improve writing. Some figures in this paper include visual elements obtained from Freepik (\url{www.freepik.com}), used under the Freepik Free License. All figure elements, including those from Freepik, are non-AI-generated. After using these tools, the author(s) reviewed and edited the content as needed and take(s) full responsibility for the content of the published article.

\bibliographystyle{elsarticle-harv} 
\bibliography{references}

\end{document}